\documentclass[12pt]{article}
\usepackage{graphics}
\usepackage{epsfig}

\newcommand{\beq}{\begin{equation}}
 \newcommand{\eeq}{\end{equation}}
\newcommand{\bea}{\begin{eqnarray}}
\newcommand{\eea}{\end{eqnarray}}

\newcommand{\eq}{\begin{equation}}
\newcommand{\eqx}{\end{equation}}
\newcommand{\eqn}{\begin{eqnarray}}
\newcommand{\eqnx}{\end{eqnarray}}
\newcommand{\f}[2]{\frac{#1}{#2}}
\newcommand{\lm}{\lambda}

\newcommand{\eps}{\varepsilon}
\newcommand{\dl}{\delta}

\renewcommand{\th}{\theta}

\newcommand{\qqqq}{\quad\quad\quad\quad}
\newcommand{\arcsinh}{\,\mbox{\rm arcsinh}\,}

\begin{document}

\begin{flushright}
{\sc\footnotesize hep-th/0510171}\\
{\sc\footnotesize NORDITA-2005-67}\\
\end{flushright}
\vspace{5pt}

\begin{center}
\vspace{24pt}
{ \large \bf Wrapping interactions and a  new source of corrections 
to the spin-chain/string duality}

\vspace{30pt}

{\sl J.\ Ambj\o rn}$\,^{a,d}$, {\sl R.\ A.\ Janik}$\,^{b}$
and {\sl C.\  Kristjansen}$\,^{c}$

\vspace{24pt}
{\footnotesize
 
$^a$~The Niels Bohr Institute, Copenhagen University\\
Blegdamsvej 17, DK-2100 Copenhagen \O , Denmark.\\
{ email: ambjorn@nbi.dk}\\
\vspace{10pt}

\vspace{10pt}
 
$^b$~Institute of Physics, Jagellonian University,\\
Reymonta 4, PL 30-059 Krakow, Poland.\\
{ email: romuald@th.if.uj.edu.pl}\\

\vspace{10pt}

$^c$~NORDITA, \\
Blegdamsvej 17, DK-2100 Copenhagen \O, Denmark.\\
{email kristjan@nbi.dk}

\vspace{10pt}

$^d$~Institute for Theoretical Physics, Utrecht University, \\
Leuvenlaan 4, NL-3584 CE Utrecht, The Netherlands.\\

\vspace{10pt}

}
\end{center}
\vspace{48pt}

\begin{center}
{\bf Abstract}
\end{center}

\noindent
Assuming that the world-sheet sigma-model  in the AdS/CFT correspondence
is an integrable {\em quantum} field theory, we deduce that there 
might be new  corrections to the spin-chain/string Bethe ansatz
paradigm. These come from virtual particles propagating around the
circumference of the cylinder and render Bethe ansatz quantization
conditions only approximate.
We determine the nature of these corrections both at weak and at strong
coupling in the near BMN limit, and find that the first
corrections behave qualitatively as wrapping interactions at weak
coupling. 

\newpage

\section{Introduction}

The discovery of integrable structures underlying planar ${\cal N}=4$ 
SYM~\cite{Minahan:2002ve,Beisert:2003tq,Beisert:2003yb,kor1} as well
as classical string theory on 
$AdS_5\times S^5$~\cite{Bena:2003wd,Arutyunov:2003uj,Dolan:2003uh,
Kazakov:2004qf,Arutyunov:2004yx,
Beisert:2005bm,Alday:2005gi} has opened up new avenues for 
testing the AdS/CFT 
correspondence~\cite{Maldacena:1997re}. Moreover, the application of spin
chain techniques has facilitated the analysis on the gauge theory
side. Non-trivial  
comparisons of gauge theory anomalous dimensions and string state energies
can now be carried out in three different regimes: The BMN 
limit~\cite{Berenstein:2002jq}, the near BMN limit~\cite{Callan:2003xr} 
and the spinning string limit~\cite{Frolov:2002av}. 
 Whereas for the first of these three
limits everything at the level of anomalous dimensions
appears to work fine the latter two are plagued by vexing
three-loop discrepancies~\cite{Callan:2004uv,Serban:2004jf}.

 It has been suggested that the three-loop discrepancy for near BMN states
and spinning strings is due to the non-commutativity of the two limits 
 which are involved in performing the analysis and which are imposed in 
different 
order on respectively the string and gauge theory side~\cite{Serban:2004jf}.
To perform an analysis on the gauge theory side in which the 
order of limits would be the same as on the string theory side one would
have to take into account 
so-called wrapping interactions~\cite{Beisert:2004hm}. 
In the spin chain language wrapping interactions are interactions 
whose range exceeds the length of the chain. It is a fact that wrapping
interactions are important when it comes to the determination of anomalous
dimensions of short operators. For instance, wrapping interactions 
contribute to the four-loop anomalous dimension of 
the Konishi operator~\cite{Beisert:2003tq,Eden:2004ua}.
A systematic
discription of wrapping interactions in terms of Feynman diagrams
can be found in~\cite{Sieg:2005kd}.
An alternative explanation of the three-loop discrepancy was suggested 
in~\cite{Minahan:2005jq}. 

 On the gauge theory side it is very hard to do any rigorous derivations 
beyond the two and three loop ones
of~\cite{Beisert:2003tq,Beisert:2003ys}, see also~\cite{kor2}. 
However, assuming
integrability and BMN scaling at higher loop 
orders a conjecture for an all loop Bethe ansatz has been put 
forward~\cite{Beisert:2004hm,Staudacher:2004tk,Beisert:2005fw}. 
On the string theory side it is possible to derive integral 
equations encoding all classical string 
solutions~\cite{Kazakov:2004qf,Kazakov:2004nh,Beisert:2005bm,Alday:2005gi}.
These integral equations can be viewed
as  classical, continuum Bethe equations. 
Inspired by these two 
achievements a suggestion for a Bethe ansatz for quantum strings 
was presented in~\cite{Arutyunov:2004vx,Staudacher:2004tk,Beisert:2005fw}.
The quantum string Bethe ansatz is identical to
the gauge theory one up to two loop order but differs from it beyond
two loops. The all loop gauge theory Bethe ansatz and the quantum string
one are very similar in nature. 
They both express the condition for factorized
scattering for a set of elementary excitations with individual momenta $p_k$
\beq
\exp(i L p_k)=\prod_{j\neq k}^M S(p_k,p_j),
\label{Bethe}
\eeq
where $S$ is the $S$-matrix for scattering of two of these excitations
and $L$ is the length of the spin chain.
The elementary excitations have the same dispersion relation on the
gauge and string theory side. What differs between the two sides is
the form of the $S$-matrices. The difference can conveniently
be encoded in a so-called dressing factor
\beq
S^{string}=\hat{S}^{dressing}S^{gauge},
\eeq
which can be expressed as a phase shift, i.e.\
\beq
\hat{S}^{dressing}(p_k,p_j)=\exp\left(i\theta(p_k,p_j)\right),
\eeq
where
\beq
\theta(p_k,p_j)=
2\sum_{r=2}^{\infty}c_r(\lambda)\left(\frac{\lambda}{16 \pi^2}\right)^r
\left(q_r(p_k)q_{r+1}(p_j)-q_{r+1}(p_k)q_r(p_j)\right),
\label{phase}
\eeq
with $\lambda$ being the 't Hooft coupling constant 
and  the $q_r$'s  certain conserved 
charges~\cite{Beisert:2004hm,Arutyunov:2004vx}. 
The expansion coefficients
$c_r(\lambda)$ must fulfill that $c_r(\lambda)\rightarrow 1$ as
$\lambda\rightarrow \infty$ in order that the classical string theory limit
is correctly reproduced.
Recently, it was found by a study of one-loop string sigma model
 corrections in the {\sl sl(2)} sector
that it is not possible to have $c_r(\lambda)=1$  for all values
of $\lambda$~\cite{Schafer-Nameki:2005tn}. The first string sigma model
loop correction produces contributions to energies of strings
spinning with total angular momentum  $L$ which
contain half integer powers of 
$\lambda'=\frac{\lambda}{L^2}$~\cite{Schafer-Nameki:2005is,Beisert:2005cw}
starting at order $(\lambda')^{5/2}$~\cite{Beisert:2005cw} as well as
non-perturbative contributions containing factors 
of the type $\exp(-\frac{1}{\sqrt{\lambda'}})$~\cite{Schafer-Nameki:2005is}.
The appearance
of terms of the type $\exp(-\frac{1}{\sqrt{\lambda'}})$
as well as terms with half-integer powers of $\lambda'$ was earlier
observed in the BMN limit in~\cite{Klebanov:2002mp} in connection with
a study of the three string interaction vertex.
The leading half integer power of $\lambda'$ in string
energies can be accounted for by
the phase factor~(\ref{phase}) if the coefficient $c_2(\lambda)$ has
the expansion~\cite{Beisert:2005cw}
\beq
c_2(\lambda)=1-\frac{3}{4}\frac{1}{\sqrt{\lambda}}
+{\cal O}\left(\frac{1}{\lambda}\right).
\eeq
This opens the interesting possibility that $c_2(\lambda)$
could tend to zero at weak coupling~\cite{Beisert:2005cw} and thus that the
three loop discrepancy for near BMN states as well as for spinning strings
could indeed be due to an order of limits problem.
Recently, it was shown that it is possible to obtain the classical 
string equations of motion of the {\sl su(2)} sector as the classical
limit of an integrable {\it quantum} field theory, namely the
$Osp(2m+2|2m)$ coset model~\cite{Mann:2005ab}. This model is defined
on the plane and not on the cylinder as needed for a quantum
string theory on $AdS_5\times S^5$. The current understanding is that
the $Osp(2m+2|2m)$ coset model is capable of capturing the quantum
effects introducing half-integer powers of $\lambda$ in the expansions
of string energies but not the finite size corrections which
appear when the theory is put on a cylinder~\cite{Mann:2005ab}.
%Of course one would still like to understand whether gauge theory 
%effects can produce a similar phase factor. As mentioned
%above, wrapping interactions have been proposed as a possible source
%of additional terms in the Bethe equations. 
%In the present paper we will

In the present paper we will explore  
the consequences of putting an integrable quantum field theory (IQFT)
on a cylinder. In particular we shall discuss what happens if the
elementary excitations of the field theory  in stead of the
standard relativistic dispersion relation obey the dispersion relation
implied by the conjectured quantum string Bethe 
equations~\cite{Arutyunov:2004vx}.
%study such wrapping interactions albeit from the point of view of the
%worldsheet of the string. 
We shall work in the near BMN limit and shall
show that wrapping interactions generically give rise to contributions
of order $\lambda^{L}$ at weak coupling and of order
$\exp(-\frac{1}{\sqrt{\lambda'}})$ at strong coupling. 

One of the established properties of an integrable {\em quantum} field
theory on a cylinder is the fact that the Bethe ansatz quantization
conditions
\eq
e^{ip_k L}=\prod_{j \neq k} S(p_k,p_j),
\eqx
are no longer exact (see e.g.\ \cite{BLZ} section 5). The effect is
very generic and comes from virtual corrections -- excitations going
around the circumference of the cylinder (see fig. 1 below). 
%As mentioned
%above the aim of this paper is to explore the consequences of 
%this effect and to
%estimate the size of these corrections both at weak and at strong
%coupling (in the near-BMN limit). 

Intuitively such processes, involving a virtual particle going around the
cylinder, when translated into Feynman graphs of the gauge theory,
should correspond to wrapping interactions. 
These types of virtual effects can be described\footnote{At leading order in
all relativistic IQFT's, and for some specific theories exactly.}
through S-matrix information, where the S-matrix is defined in the
infinite volume 
limit. Thus, since there are proposals for the asymptotic S-matrix for
gauge theory/strings we propose to interpret them as the infinite
volume scattering data and use the framework of the virtual
corrections to IQFT to incorporate the effect of wrapping
interactions. We show that at weak coupling the order at which the
virtual corrections set in is $\lambda^L$ just as
expected for wrapping interactions.

The outline of this paper is as follows. In section 2 we will review
these corrections in ordinary relativistic integrable field
theories. In section 3 we will motivate what changes need to be done
in order to incorporate the nonstandard dispersion relation characteristic 
of excitations in the conjectured long range Bethe ansatz, then in section 4 we
estimate the size of the corrections both at weak coupling and in the
near-BMN limit. We close the paper with a summary and outlook.

\section{Finite size mass shift in relativistic integrable field
  theories}

In \cite{lu1,lu2} L\"{u}scher calculated the leading order corrections to
the (infinite volume) masses/energies of 1-particle states when put on a
cylinder of circumference $L$:
\eq
m(L)=m(L=\infty)+\Delta m_\mu (L)+\Delta m_F (L).
\eqx
These corrections arise from two different types of
processes: the first is the so-called $\mu$-term which arises when the
particle splits into a pair of virtual on-shell particles which go
around the cylinder and recombine later, while the second one, the
$F$-term comes from a virtual particle loop where the virtual particle
goes around the circumference of the cylinder (see fig. 1). 
\begin{figure}[htb]
%\vspace{-3cm}
\centerline{\scalebox{0.5}{\rotatebox{0}{\includegraphics{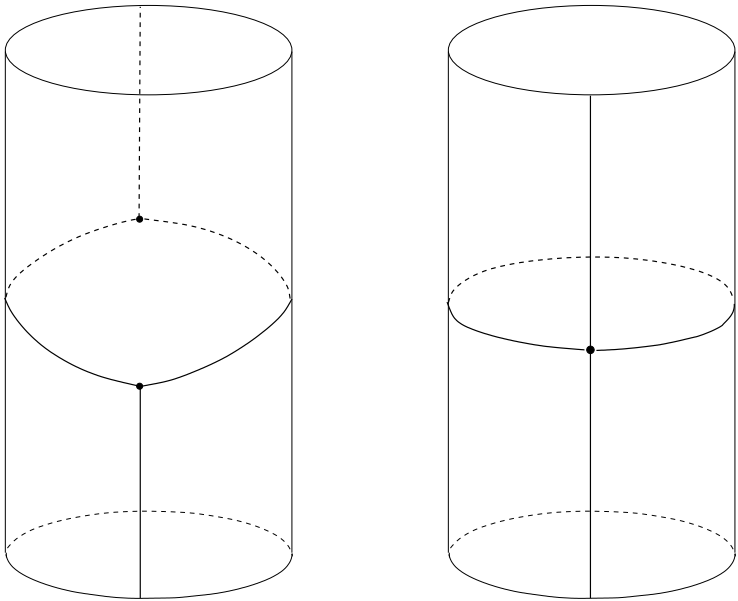}}}}
%\vspace{-3.5cm}
\caption[fig1]{The diagram to the left (the $\mu$-term) shows a particle
splitting in two virtual, on-shell particles, traveling around the 
cylinder and recombining. The diagram to the right (the F-term)
shows a virtual particle going around the circumference of the cylinder.}
\label{fig1}
\end{figure}

The
expressions for these terms in the case of 1+1 dimensions and a single
mass scale in the theory are \cite{km2}
\eqn
\label{e.muterm}
\f{\Delta m_\mu (L)}{m(\infty)} &=&  -\f{\sqrt{3}}{2} \sum_{b,c} M_{abc} (-i)
\,{\mbox{\rm res}}_{\th=2\pi i/3}\, S_{ab}^{ab}(\th) \cdot
e^{-\f{\sqrt{3}}{2}m L}, \\ 
\label{e.fterm}
\f{\Delta m_F (L)}{m(\infty)} &=& -\int_{-\infty}^\infty
\f{d\th}{2\pi} e^{-m L \cosh \th} \cosh \th \sum_b
\left(S_{ab}^{ab}\left(\th+i \f{\pi}{2} \right)-1 \right). 
\eqnx
where $S_{ab}^{ab}(\th)$ is the (infinite volume) $S$-matrix, and
$M_{abc}=1$ if $c$ is a bound state of $a$ and $b$ and zero otherwise.
These formulas have been checked to agree with a wide variety of
examples such as Ising field theory, the so-called scaling Lee-Yang model
(SLYM), various perturbed minimal model CFT's etc. \cite{km1,km2}.

However, it is rather difficult to use these formulas directly in the
case of the world sheet integrable field theories appearing in our
present context. The proofs of L\"{u}scher's formulas were a mixture of
diagrammatic analysis and analytical continuation where Lorentz
invariance was crucial (although clearly 
many aspects remain valid on the lattice \cite{munster}). 
In particular, one always had the dispersion
relation $E^2-p^2=m^2$ and the parametrization of energy and momentum
through rapidity
\eq
E=m\cosh \th, \qqqq p=m \sinh \th.
\eqx
For the world sheet string theories in {\em curved} AdS space-time it
was pointed out \cite{Lightcone,Callan:2003xr,Callan:2004uv,Arutyunov:2004yx}
that the usual light cone gauge like 
$\tau=t$ (where $t$ is e.g.\ the global AdS time) is inconsistent with
putting a Minkowskian metric on the world sheet. Hence the second gauge
condition is usually taken to be the uniform gauge where the density
of a conserved charge (such as the R-charge $J$) is spread uniformly
along the string\footnote{Usually one makes the condition $p_\phi=J$
  keeping the size of the cylinder to $2\pi$ \cite{Arutyunov:2004yx}. But this
  populates the action with explicit factors of $J$. It is more
  convenient for our purposes to set the density to be equal to 1 and
  then the total R-charge is encoded just in the size of the cylinder like in
  \cite{gleb2}.}. Then the gauge-fixed field theory is {\em not}
explicitly Lorentz-invariant! 
A common ingredient of the asymptotic Bethe ans\"{a}tze is the dispersion
relation for all elementary excitations (here $E\equiv \Delta-J$):
\eq
\label{e.dispersion}
E=\sqrt{1+8g^2 \sin^2\f{p}{2}},\hspace{1.2cm}g^2=\frac{\lambda}{8\pi^2}.
\eqx    
In the following we will {\em assume} that this is indeed the true
dispersion relation for excitations in the integrable quantum field
theories on the world sheet and explore the consequences of this
assumption. 

Another point which one should mention is that e.g.\ in the near BMN
limit the zero momentum 1-particle states are protected by
supersymmetry and what one is really interested in are corrections to
2-particle (or higher) states for which we lack similar expressions.

In order to attack the above problems it is convenient to look at a
way in which the finite size virtual corrections can be calculated
exactly, and which may shed more light on the origin of the formulas
(\ref{e.muterm}) and (\ref{e.fterm}). 
 
There are basically three (interrelated) ways of computing exactly the
spectrum of an integrable QFT on a cylinder: the Thermodynamic Bethe
Ansatz (TBA) \cite{zamTBA}, Nonlinear Integral equation (NLIE) or
Destri-de-Vega (DDV) equation \cite{DDV}, and functional relations
\cite{BLZ}. All this is still an area of active research \cite{xxxx}.
We will here concentrate on the first of these approaches. 

Before we start let us emphasize that the Thermodynamic Bethe Ansatz
that we consider here is {\em not} the thermodynamic approximation
which is used for approximately solving Algebraic Bethe Ansatz
equations for a large number of roots in the spin chain/string
literature. We use the term in the sense that it is used in the context
of relativistic integrable field theories.  

\subsubsection*{Thermodynamic Bethe Ansatz}

The Thermodynamical Bethe Ansatz \cite{zamTBA} was originally devised
as a method for finding how the ground state of an integrable field
theory depends 
on the circumference $L$. The main idea is to consider
the theory on a very 
elongated torus with circumference $L$ and length $R \to \infty$, and to
compute the (euclidean) partition function. From the point of view of
$L$ being space and $R$ being very large `time' this gives the ground state
energy being calculated:
\eq
E_0(L)=-\f{1}{R} \log Z.
\eqx    
On the other hand if one looks at $R$ being space and $L$ being the
(euclidean) time this is just the thermal free energy of the system
on an {\em infinite} line, which can be calculated by
minimising $E-TS$ with $T=1/L$. The main point is that for the theory
with $R\to \infty$, Bethe ansatz quantization remains exact. Then one
introduces continuous densities for the occupied roots AND appropriate entropy
factors. The result is the set of equations\footnote{For the simplest
  case of a single particle in the spectrum.}:
\eqn
\label{e.eps}
\eps(\th) &=& L E_{TBA}(\th) -\phi * L, \\
\label{e.en}
E_0 &=& \int_{-\infty}^\infty \f{d\th}{2\pi} p'_{TBA}(\th) L(\th), 
\eqnx  
where
\eqn
L(\th) &=& \log(1+e^{-\eps(\th)}), \\
(\phi * L)(\th) &=& \int_{-\infty}^\infty \f{d\th'}{2\pi}
\phi(\th-\th') L(\th'), \\
\phi(\th) &=& -i \f{d}{d\th} \log S(\th),
\eqnx
and $E_{TBA}(\th)=m \cosh \th$, and $p_{TBA}(\th)=m\sinh \th$.

As it stands the Thermodynamic Bethe Ansatz seems to be confined to
being a tool for finding just the ground state energy. However, in
\cite{DoreyTateo1} Dorey and Tateo  suggested that by analytical continuation
one could obtain the energies of excited states. This program was
carried out first in the scaling Lee-Yang model 
(SLYM) \cite{DoreyTateo1} and then in a series of
perturbed minimal model CFT's \cite{DoreyTateo2}.

The underlying mechanism of the construction is that
$L(\th)=\log(1+e^{-\eps(\th)})$ developed singularities which after a contour
deformation added source terms to the right hand sides of equations
(\ref{e.eps}) and (\ref{e.en}). E.g. for a 1-particle state at zero
momentum, the $mL \cosh \th$ in (\ref{e.eps})
should be substituted by
\eq
mL \cosh \th +\log\f{S(\th-\th_0)}{S(\th+\th_0)},
\eqx
where $\th_0$ is a singularity of $L(\th)$, which leads to an
additional equation $\eps(\th_0)=i\pi$. This set of
equations then gives {\em exact} results for the excited state energy
$E_1(L)$ for any $L$. Moreover, 
L\"{u}scher's formulas for (\ref{e.muterm}) and (\ref{e.fterm}) then
immediately follow from solving that system of equations by iteration
(see \cite{DoreyTateo1}).

For our purposes it is interesting to obtain the form of corrections
for a 2-particle state for which we lacked an explicit expression. 

From the results in \cite{DoreyTateo1} we obtain 
\eqn
\label{e.2part}
E&=&2\cosh \th_B +2\cosh \th_B \cdot \dl -\nonumber\\ 
&&\int_{-\infty}^\infty
\f{d\th}{2\pi} \cosh \th e^{-L\cosh\th} S\left(\th+i\f{\pi}{2}-\th_B
\right) S\left(\th+i\f{\pi}{2}+\th_B \right), 
\eqnx 
where $\th_B$ is the Bethe root which follows from solving the
Bethe quantization condition
\eq
e^{iL \sinh \th_B}=S(2\th_B).
\eqx
We see that the first piece of (\ref{e.2part}) is the classical energy
of the 2-particle state. The second piece is a $\mu$-term of the form
\eq
\label{e.muterm2}
\dl=3 e^{-\f{\sqrt{3}}{2} L \cosh \th_B} \coth \th_B \sqrt{ \f{2\cosh
    2\th_B+1}{2\cosh 2\th_B-1} },
\eqx  
while the last term is the analogue of the $F$-term. The exponential
factor remains the same, while the two S-matrices correspond to a
virtual particle of rapidity $\th+i \pi/2$ moving around the cylinder
and scattering off the two particles with Bethe roots (rapidities) $\pm
\th_B$. 

The point that we want to make here is that the structure of the
$F$-term correction to the 2-particle state is a straightforward
generalization of the L\"{u}scher mass shift formula for a 1-particle
state. In view of applications to the world sheet theory we would
expect a first nontrivial correction to occur at the level of
2-particle states since the mass of the 
one-particle states are protected by supersymmetry. 
The absence of corrections for the ground state and
1-particle states should be a property of the full world sheet theory
and cannot be obtained just in e.g.\ the {\sl su(2)} sector. The reason
is that the   {\sl su(2)} excitation scatters non-trivially from other
excitations \cite{Beisert:2005fw} which might therefore circulate in the loop.

Another motivation for discussing at length the TBA derivation of L\"{u}scher's
formulas is that the space-time interpretation of the derivation of TBA
may suggest its generalization to the case of the world sheet theory
which, when gauge-fixed, is no longer Lorentz-invariant.

In the next section we will  use the TBA intuition to propose
what changes are needed when the dispersion relation is no longer exactly
relativistic but is modified to (\ref{e.dispersion}). We will also try
to motivate, using a solvable example, that such a procedure can be
legitimate -- which is of course not completely clear {\em a-priori}.

\section{Nonstandard dispersion relations}

As we saw above, the derivation of the TBA is based on a modular
transformation on the torus i.e. an interchange of space and time. 
For a relativistic theory this does not really make a difference but it is
far from clear if such a philosophy may be applied at all to the
rather nonstandard world-sheet theories.

In order to motivate this we will show an example where the proposed
procedure can be proved to give the exact result despite the fact that
the dispersion relation is not exactly relativistic.

\subsubsection*{The Ising model off criticality}

This example has been given by L\"{u}scher in his Cargese lectures
\cite{lu1}. We need to be here more explicit in the
calculation, as the intermediate steps will be crucial for our
purposes. The Ising model
on a finite $L\times L$ lattice can be completely solved in terms of
transfer matrices which can be expressed through free fermion
operators with the dispersion relation \cite{SML}:
\eq
\label{e.dising}
\cosh E_q= \cosh m_0+1-\cos q \equiv \cosh m_0+2\sin^2 \f{q}{2}.
\eqx
The mass gap (i.e. the mass/energy of the lowest lying excitation) can be
expressed through an exact formula:
\eq
\label{e.is1}
m(L)=m_0+\f{1}{2} \sum_{\nu=0}^{L-1} E_{\f{\pi}{L}(2\nu+1)} -\f{1}{2}
\sum_{\nu=0}^{L-1} E_{\f{\pi}{L}\cdot 2\nu} =m_0-\f{1}{2}
\sum_{\nu=0}^{2L-1} e^{i\pi \nu} E_{\f{\pi}{L}\cdot \nu}.
\eqx 
The discrete sum can be rewritten using the (finite) Poisson resummation
formula \cite{crowley} as
\eq
\label{e.is2}
m(L)=m_0-L \sum_{\nu=-\infty}^\infty \int_{-\pi}^\pi \f{dq}{2\pi} e^{i
  qL(2\nu+1)} E_q.
\eqx 
Now after an integration by parts one changes variables to
\eq
Q=-i E_q,
\eqx
$Q$ will play the role of the new momentum in what follows.
For {\em integer} $L$ one then gets
\eq
m(L)=m_0+\sum_{\nu=-\infty}^\infty \f{1}{2\nu+1}
\int_{-iE_{-\pi}}^{-iE_{\pi}} \f{dQ}{2\pi} e^{i q(Q) L (2\nu+1)}.
\eqx
We note that due to the form of (\ref{e.dising}) $q(Q)=i E_Q$ so the
exponent simplifies to $-E_Q L (2\nu+1)$. 
After a slightly nontrivial contour argument  it
turns out that one can change the contour of integration to go from
$-\pi$ to $\pi$ just as if $Q$ was a physical momentum. One then gets
\eq
\label{e.isingfin}
m(L)=m_0+\sum_{\nu=0}^\infty \f{2}{2\nu+1} \int_{-\pi}^\pi
\f{dQ}{2\pi} e^{-L (2\nu+1) E_Q} \sim m_0+2\int_{-\pi}^\pi
\f{dQ}{2\pi} e^{-L E_Q}
\eqx 
The leading correction is then the term with $\nu=0$. The exponent is
just as in the formula for relativistic $F$-term correction (with the S matrix
equal to $-1$, which is the value for the massive deformation of 
the Ising model). 

\subsubsection*{Space-time interchange}

Let us now go back and analyze what kind of analytical
continuations were made here and adopt a notation more reminiscent of
the TBA expressions. What plays the role of momentum in the
final formula (\ref{e.isingfin}) is 
\eq
\label{e.ptba}
p_{TBA}\equiv Q = -i E_q,
\eqx
while the energy appearing in the exponent of (\ref{e.isingfin}) is in
fact $-i$ times the original momentum expressed in terms of $Q\equiv p_{TBA}$:
\eq
\label{e.etba}
E_{TBA}=-i q(Q).
\eqx 

The above substitutions are indeed quite natural from the point of view of
a modular transformation exchanging space and time. Since time
should be interchanged with space one should Wick rotate both
coordinates (hence the $-i$'s in  (\ref{e.ptba}) and (\ref{e.etba})),
and also exchange $E$ and $q$ with each other. In the final integral
(\ref{e.isingfin}) $p_{TBA}\equiv Q$ is taken to be real.

To add some more plausibility to the interpretation of the above
formulas, one sees that in the Lorentz invariant case the dispersion
relation in the double Wick rotated space remains the same and the
shift of the rapidity of the virtual particle by $i\pi/2$ in the
$F$-term exactly corresponds to the above transformations
(\ref{e.ptba})--(\ref{e.etba}).
\eq
m\cosh(\th+i\pi/2)=i m \sinh\th, \qqqq
m\sinh(\th+i\pi/2)=i m \cosh\th.
\eqx

In the following section we will apply these ideas to estimate the size of
the correction terms for the world sheet theory with the dispersion relation
(\ref{e.dispersion}) but before we do that, we will show that the above
procedure works for a modified Ising model with (\ref{e.dispersion}).

\subsubsection*{Modified Ising model}

Since we lack any explicit example of a completely defined integrable
theory with the dispersion relation (\ref{e.dispersion}), we will
construct a somewhat artificial but exactly solvable example. Let us
modify the Ising model defined by the transfer matrices involving free
fermions by changing their dispersion relation from (\ref{e.dising}) to
(\ref{e.dispersion}).

The exact mass shift can then be obtained in exactly the same
fashion. It is given by formulae analogous to
(\ref{e.is1})-(\ref{e.is2}), i.e.
\eq
m(L)=1-L \sum_{\nu=-\infty}^\infty \int_{-\pi}^\pi \f{dq}{2\pi} e^{i
  qL(2\nu+1)} \sqrt{1+8g^2 \sin^2\f{q}{2}}.
\eqx
Let us now evaluate $E_{TBA}(p_{TBA})$. Performing the substitutions
(\ref{e.ptba})-(\ref{e.etba}) we obtain
\eq
\label{e.newe}
E_{TBA}(p_{TBA})=2\arcsinh \left( \f{1}{2\sqrt{2}g} \cdot
  \sqrt{1+p_{TBA}^2} \right). 
\eqx
Now again the quite remarkable identity holds for integer $L$:
\bea
\lefteqn{
-L \int_{-\pi}^\pi \f{dq}{2\pi} e^{i qL(2\nu+1)} \sqrt{1+8g^2
   \sin^2\f{q}{2}} } \nonumber \\
 && \hspace*{2.0cm}=\f{1}{2\nu+1} \int_{-\infty}^\infty\f{dq}{2\pi}
e^{-L(2\nu+1) 
  \cdot 2\arcsinh \left( \f{\sqrt{1+p_{TBA}^2}}{2\sqrt{2}g} 
  \right)}, 
\eea
which for $\nu=0$ reproduces the expected correction
\eq
m(L)=1+2 \int_{-\infty}^\infty\f{dq}{2\pi}
e^{-L \cdot 2\arcsinh \left( \f{\sqrt{1+p_{TBA}^2}}{2\sqrt{2}g} 
  \right)}
\eqx
but with the somewhat odd-looking expression (\ref{e.newe}) for the energy.

\section{Wrapping interactions and near-BMN limit}

Let us now apply the above considerations to the case of the
world sheet theory with the dispersion relation
(\ref{e.dispersion}). We will evaluate at what order does the virtual
correction enter both at weak and at strong gauge theory coupling (in
the near BMN limit). We will not try here to obtain the exact form of the
correction since this would have to involve the full world sheet theory
whose S-matrix is so far not known completely. Moreover the scattering is
non-diagonal making the procedure even more complicated. The
conjectured Bethe ans\"{a}tze \cite{Beisert:2005fw} may be 
used in this respect but
carrying out this program still remains a very nontrivial task which
we leave for future investigation. 

In the following we will discuss first the $F$-term and then the $\mu$-term.

\subsubsection*{The $F$-term}

The size of the $F$ term at large $L$ is governed by the exponential
factor
\eq
\label{e.tbaexp}
e^{-L E_{TBA}} =e^{-L\cdot  2\arcsinh \left(
    \f{\sqrt{1+p_{TBA}^2}}{2\sqrt{2}g}  \right)}.
\eqx
Let us see how  this expression behaves at small coupling. Then the
argument of the $\arcsinh$ is very large and we may substitute it by a
logarithm $E_{TBA} \sim -2 \log g+\ldots$. Consequently the correction term is
of the order
\eq
g^{2L},
\eqx
which is exactly the expected order when wrapping interactions should
set in. This further supports the intuition that virtual corrections
in the world sheet theory corresponding to particles moving in loops
around the cylinder should correspond to wrapping interactions on the
gauge theory side.

Let us now proceed to take the near BMN limit. We take\footnote{For a
  two impurity state in the {\sl su(2)} sector $L=J+2$.} $L\sim J\to \infty$
but at the same time keep $\lm'=\lm/J^2$. Therefore
\eq
8g^2=\f{\lm' J^2}{\pi^2}.
\eqx
In this limit the argument of the $\arcsinh$ is small and the exponent
  (\ref{e.tbaexp}) governing the size of the corrections takes the
  form\footnote{We note that in this limit the same result would also
  arise from an unmodified BMN dispersion relation $E=\sqrt{1+2g^2 p^2}$.}
\eq
e^{-\f{2\pi}{\sqrt{\lm'}} \sqrt{1+p_{TBA}^2}}.
\eqx
%So while the virtual corrections are not of the correct 
%form to explain in a simple way 
%the 3-loop discrepancy between gauge and string theory,
%it is nevertheless of the form  mentioned in the 
%introduction and could play an important role
%resolving the discrepancy.

\subsubsection*{$\mu$-term}

We have much less control over the precise form of the $\mu$-term,
however its physical origin is quite clear. It occurs when there is a
possible process when a particle can decay into a pair of virtual
particles which are {\em on-shell} (but yet will have in general complex
momenta).

For  Lorentz-invariant theories, 
the exponents of the $\mu$-terms (\ref{e.muterm}) and
(\ref{e.muterm2}) are seen to follow from the imaginary
parts of the momenta of the two virtual particles. Explicitly, suppose
that an on-shell 
particle of momentum $p_0$ disintegrates into a pair of particles with
momenta $p_1$ and $p_2$ which are on-shell. From the kinematics
we obtain for $p_{1,2}$:
\eq
p_{1,2}=\f{p_0}{2} \pm i\f{\sqrt{3}}{2} \sqrt{1+p_0^2}.
\eqx
We do not have a precise expression for the $\mu$-term in the case of
multi-particle states. We will here just use an ad-hoc formula
$\exp(-L\cdot  Im\, p_i)$, and we find that it 
reproduces the exponents in (\ref{e.muterm}) and
(\ref{e.muterm2}).

Repeating the same kind of calculation with the dispersion relation
(\ref{e.dispersion}) gives in general quite complicated
expressions. We may however take $p_0$ to be the momentum of a
particle in a 2-particle state in the near BMN limit:
\eq
p_0=\f{2\pi n}{J+2}+\f{2\pi n}{\sqrt{1+\lm' n^2} (J+2)^2} +\ldots
\eqx 
Assuming $p_i$ to be small we obtain
\eq
p_{1,2}=\f{1}{J} \left(n \pi \pm i \f{\sqrt{3}}{\sqrt{\lm'}} \pi
  \sqrt{1+\lm' n^2} \right).
\eqx
This suggests that the $\mu$-term, if present, would also give
$1/\sqrt{\lm'}$ corrections in the near BMN limit.

\section{Conclusions}

%%%%

Lacking a direct proof of the AdS/CFT correspondence, tests which 
are truly reflecting non-trivial interactions  
are of utmost importance. The integrable structures
discovered both at the string side and the ${\cal N}=4$ SYM side
provide a good testing ground and there has been extensive research
comparing the predictions coming from the string world-sheet sigma
model with the predictions from the spin-chain model of ${\cal N}=4$
SYM. 

Numerous examples of very detailed agreement between gauge and
string theory have been found, 
nevertheless there remains some subtle but significant
discrepancies. One notable example is the notorious three loop
discrepancy occurring both in the near-BMN limit and for spinning
strings. Another is the appearance half-integer powers of $\lambda$
in string sigma model loop corrections, cf.\ the introduction.
However, it is known that at least some ingredients have
not been included in the asymptotic Bethe ans\"{a}tze such as
e.g. gauge theory wrapping interactions.  

%%%

While integrability on the string theory side is fairly developed for
the classical sigma model 
not much is known for the complete quantum theory. A very recent 
attempt to study the quantum effects via S-matrix (Bethe Ansatz) 
techniques can be found in the paper by Mann and Polchinski 
\cite{Mann:2005ab}, who considered directly a relativistic quantum
field theory on $R^2$ (a $OSp(2m+2|2m)$ supercoset model) with a known
S-matrix. They found an embedding of the `classical' string {\sl su(2)}
sector into the Bethe ansatz of their full quantum theory,
and found generic corrections of the type $1/\sqrt{\lambda}$, 
but did not consider finite size effects.

In our paper we wanted to concentrate on the specific features of an
integrable {\em quantum} field theory when put on a cylinder, as is
always ultimately required for the closed string worldsheet
theory. The generic feature is then that the Bethe ansatz quantization
condition is no longer {\em exact} and receives virtual corrections
from particles moving in loops around the circumference of the
cylinder.

The energy shifts due to these effects in a finite volume are most 
conveniently calculated using the Thermodynamic Bethe
Ansatz.\footnote{Although no  general proof of this method for excited
  states exists, extensive
  tests have been performed in various relativistic integrable field
  theories~\cite{DoreyTateo1,DoreyTateo2,Balog}.}  
We provided arguments
that despite the non-Lorentz-invariant nature of the dispersion relation
(\ref{e.dispersion}) conjectured to hold for 
the excitations of the $AdS$ string, one can
still use TBA. It does not seem possible at this stage to perform a
complete calculation for the world-sheet theory. Therefore,
in this paper we have confined ourselves to understand
just the general nature of the corrections, using only the 
dispersion relation (\ref{e.dispersion}) as an assumption.

%%%

%In principle one could 
%use the conjectured Bethe Ans\"{a}tze in \cite{Beisert:2005fw} as a
%starting point for a full analysis, however it would be a very
%formidable task not only technically but also conceptually. 
%Therefore we have confined ourselves to understand just the 
%general nature of the corrections, using only the 
%dispersion relation (\ref{e.dispersion}) as an assumption. 

%%%

At weak coupling we find that the first 
corrections are of the order $\lambda^{L}$. It is natural 
to conjecture that the analogous term on the gauge-theory side 
is due to wrapping interactions in the spin-chain picture. This is
also very intuitive as the propagation of a virtual excitation around
the cylinder would presumably, when translated somehow into 
gauge theory Feynman
graphs, be described by graphs `wrapping' the cylinder.

At strong coupling the contribution is of the order $e^{-1/\sqrt{\lambda'}}$
which cannot be seen directly in perturbative gauge theory. 
However, as mentioned earlier, in string sigma model loop corrections
one has found contributions to string energies involving half
integer powers of $\lambda'$ as well as contributions of
of the form $e^{-1/\sqrt{\lambda'}}$
~\cite{Schafer-Nameki:2005tn,Schafer-Nameki:2005is,Beisert:2005cw},
see also~\cite{Klebanov:2002mp}.

It would be very interesting to perform a more complete calculation,
perhaps using all or a part of the asymptotic Bethe ans\"{a}tze of
\cite{Beisert:2005fw}. The same could be also considered in the
supercoset model of \cite{Mann:2005ab}, which has the additional
simplification that it is a true relativistic integrable field theory
-- so it does not suffer from the conceptual difficulties associated
with the nonstandard dispersion relation. Last but not least, it would
be very interesting to try to extend the formalism of such finite size
virtual corrections to many particle states dual to macroscopic
spinning strings.

\bigskip

\noindent{\bf Acknowledgments.} The authors were all supported by
ENRAGE (European Network on Random Geometry), a Marie Curie
Research Training Network financed by the European Community's
Sixth Framework Programme, network contract MRTN-CT-2004-005616.
RJ thanks the NBI for hospitality when this work was carried out.
RJ was supported in part by Polish Ministry
of Science and Information Society Technologies  
grants 1P03B02427 (2004-2007) and 1P03B04029 (2005-2008).

\end{document}